\documentclass[showpacs,twocolumn,floatfix,superscriptaddress]{revtex4}
\usepackage{graphicx}
\usepackage{amsmath}
\begin{document}
\title{Distribution of voltage fluctuations in a current-biased conductor}
\author{M. Kindermann}
\affiliation{Instituut-Lorentz, Universiteit Leiden, P.O. Box 9506, 2300 RA
Leiden, The Netherlands}
\author{Yu.\ V. Nazarov}
\affiliation{Department of Nanoscience, Delft University of Technology,
Lorentzweg 1, 2628 CJ Delft, The Netherlands}
\author{C. W. J. Beenakker}
\affiliation{Instituut-Lorentz, Universiteit Leiden, P.O. Box 9506, 2300 RA
Leiden, The Netherlands}
\date{October 2002}
\begin{abstract}
We calculate the fluctuating voltage $V(t)$ over a conductor driven out of
equilibrium by a current source. This is the dual of the shot noise problem of
current fluctuations $I(t)$ in a voltage-biased circuit. In the single-channel
case the distribution of the accumulated phase $\Phi=(e/\hbar)\int Vdt$ is the
Pascal (or binomial waiting-time) distribution --- distinct from the binomial
distribution of transferred charge $Q=\int Idt$. The weak-coupling limit of a
Poissonian $P(\Phi)$ is reached in the limit of a ballistic conductor, while in
the tunneling limit $P(\Phi)$ has the chi-square form.
\end{abstract}
\maketitle

The current--voltage or charge--phase duality plays a central role in the
theory of single-electron tunneling through tunnel junctions of small
capacitance \cite{ASI}. At the two extremes one has a voltage-biased junction
(in which the voltage is kept fixed by a source with zero internal resistance,
while the current fluctuates) and a current-biased junction (fixed current from
a source with infinite internal resistance, fluctuating voltage). In the
current-biased case the Coulomb blockade induces periodic oscillations in the
voltage \cite{Ave91}, while in the voltage-biased case the Coulomb blockade is
inoperative.

Quantum mechanically, the duality appears because current $I$ and voltage $V$
are noncommuting operators \cite{Sch90}. This is conveniently expressed by the
canonical commutator $[\Phi,Q]=ie$ of the transferred charge
$Q=\int_{0}^{\tau}I(t)dt$ and accumulated phase
$\Phi=(e/\hbar)\int_{0}^{\tau}V(t)dt$ (in a given detection time $\tau$).
Moments of charge and phase determine the measured correlators of current and
voltage, respectively \cite{note1}.

While all moments of $Q$ in a voltage-biased conductor are known
(\cite{Lev93}),
the dual problem (moments of $\Phi$ under current bias) has only been studied
for the first two moments \cite{Ben83,Lee96}. In the absence of
Coulomb-blockade
effects, the first two moments in the dual problems are simply related by
rescaling $I(t)\rightarrow V(t)\times G$ (with $G$ the conductance). One might
surmise that this linear rescaling carries over to higher moments, so that the
dual problems are trivially related in the absence of the Coulomb blockade.
However, the rescaling (as derived for example in Ref.\ \cite{Bla00})
follows from a Langevin approach that is suspect for moments higher than the
second --- so that one might expect a more complex duality relation.

The resolution of this issue is particularly urgent in view of recent proposals
to measure the third cumulant of shot noise in a mesoscopic conductor
\cite{Lev01,Gut02,Nag02}. Does it matter if the circuit is voltage biased or
current biased, or can one relate one circuit to the other by a linear
rescaling? That is the question addressed in this paper.

We will demonstrate that, quite generally, the rescaling breaks down beyond the
second moment. We calculate all moments of the phase (hence all correlators of
the voltage) for the simplest case of a single-channel \cite{spin} conductor
(transmission probability $\Gamma$) in the (zero-temperature) shot-noise limit.
In this case the charge $Q\equiv qe$ for voltage bias $V_{0}\equiv
h\phi_{0}/e\tau$ is known to have the binomial distribution \cite{Lev93}
\begin{equation}
P_{\phi_{0}}(q)={\phi_{0} \choose q}\Gamma^{q}(1-\Gamma)^{\phi_{0}-q}.
\label{Pqresult}
\end{equation}
We find that the dual distribution of phase $\Phi\equiv 2\pi\phi$ for current
bias $I_{0}\equiv eq_{0}/\tau$ is the Pascal distribution \cite{note2}
\begin{equation}
P_{q_{0}}(\phi)={\phi-1 \choose q_{0}-1}\Gamma^{q_{0}}
(1-\Gamma)^{\phi-q_{0}}.\label{Pphiresult}
\end{equation}
(Both $q$ and $\phi$ are integers for integer $\phi_{0}$ and $q_{0}$.)

In the more general case we have found that the distributions of charge and
phase are related in a remarkably simple fashion for $q,\phi\rightarrow\infty$:
\begin{equation}
\ln P_{q}(\phi) = \ln P_{\phi}(q) + {\cal O}(1).\label{PCP}
\end{equation}
(The remainder ${\cal O}(1)$ equals $\ln(q/\phi)$ in the shot-noise limit.)
This manifestation of charge-phase duality, valid with logarithmic accuracy,
holds for any number of channels and any model of the conductor.

We have obtained these results in the course of an extensive study of the
quantum statistics of fluctuating currents and voltages in electrical circuits
containing mesoscopic conductors. Our method is the non-equilibrium Keldysh
action technique. Within this technique, the electronic degrees of freedom are
traced out and the quantum dynamics is represented by path integrals over
fluctuating phases on the Keldysh contour \cite{Sch90}. Before presenting the
derivation we give an intuitive physical interpretation.

The binomial distribution (\ref{Pqresult}) for
voltage bias has the interpretation \cite{Lev93} that electrons hit the barrier
with frequency $eV_{0}/h$ and are transmitted independently with probability
$\Gamma$. For current bias the transmission rate is fixed at $I_{0}/e$.
Deviations due to the probabilistic nature of the transmission process are
compensated for by an adjustment of the voltage drop over the barrier. If the
transmission rate is too low, the voltage $V(t)$ rises so that electrons hit
the barrier with higher frequency. The number of transmission attempts
(``trials'') in a time $\tau$ is given by $(e/h)\int_{0}^{\tau}V(t)dt\equiv
\phi$. The statistics of the accumulated phase $\phi$ is therefore given by the
statistics of the number of trials needed for $I_{0}\tau/e$ successful
transmission events. This stochastic process has the Pascal distribution
(\ref{Pphiresult}).

Starting point of our derivation is a generalization to time-dependent bias
voltage $V(t)=(\hbar/e)\dot\Phi(t)$ of an expression in the literature
\cite{Lev93,Naz99} for the generating functional ${\cal Z}[\Phi(t),\chi(t)]$ of
current fluctuations:
\begin{eqnarray}
&&{\cal Z}[\Phi,\chi]=\left\langle\overleftarrow{\rm
T}\exp\left\{\frac{i}{e}\int
dt\bigl[\Phi(t)+{\textstyle\frac{1}{2}}\chi(t)\bigr]\hat{I}(t)\right\}\right.
\nonumber\\
&&\;\;\;\mbox{}\left.\times\overrightarrow{\rm T}\exp\left\{\frac{i}{e}\int
dt\bigl[-\Phi(t)+{\textstyle\frac{1}{2}}\chi(t)\bigr]
\hat{I}(t)\right\}\right\rangle.\label{Zphichi}
\end{eqnarray}
(The notation $\overrightarrow{T}(\overleftarrow{T})$ denotes time-ordering of
the exponentials in ascending (descending) order.) Functional derivatives of
the Keldysh action $\ln{\cal Z}$ with respect to $\chi(t)/e$ produce
irreducible correlators of the current operator $\hat{I}(t)$ to any order
desired. This expression provides an example of the ``tracing out'' of degrees
of freedom; The average is over electronic degrees of freedom, the electrical
current being the only operator coupled to the phase.

\begin{figure}
\includegraphics[width=6cm]{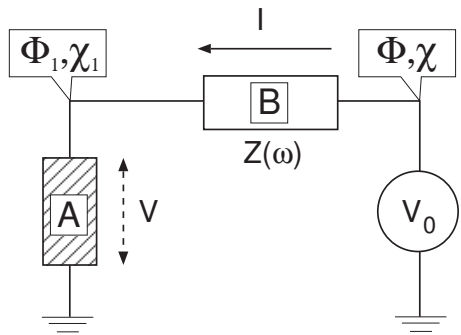}
\caption{
Mesoscopic conductor (shaded) in a circuit containing a voltage source $V_{0}$
and series impedance $Z(\omega)$. Both the current $I$ through the circuit and
the voltage drop $V$ over the conductor may fluctuate in time. The dual
problems contrasted here are: voltage bias ($Z\rightarrow 0$, fixed $V=V_{0}$,
fluctuating $I$) and current bias ($Z\rightarrow \infty$, fixed $I=V_{0}/Z$,
fluctuating $V$). The phases $\Phi,\chi$ appearing in Eq.\
(\protect\ref{concatenation}) are indicated.
\label{circuit}
}
\end{figure}

To make the transition from voltage to current bias we introduce a second
conductor $B$ in series with the mesoscopic conductor $A$ (see Fig.\
\ref{circuit}). The generating functional ${\cal Z}_{A+B}$ of current
fluctuations in the circuit is a (path integral) convolution of ${\cal Z}_{A}$
and ${\cal Z}_{B}$,
\begin{equation}
{\cal Z}_{A+B}[\Phi,\chi]=\int{\cal D}\Phi_{1}{\cal D}\chi_{1}\,{\cal
Z}_{A}[\Phi_1,\chi_1]{\cal
Z}_{B}[\Phi-\Phi_{1},\chi-\chi_{1}].\label{concatenation}
\end{equation}
One can understand this expression as the average over fluctuating phases
$\Phi_{1},\chi_{1}$ at the node of the circuit shared by both conductors.

In general the functional dependence of ${\cal Z}_{A},{\cal Z}_{B}$ is rather
complicated and non-local in time, but we have found an interesting and
tractable low-frequency regime: The non-locality may be disregarded for
sufficiently slow realizations of the fluctuating phases. In this regime the
functional ${\cal Z}$ can be expressed in terms of a function $S$,
\begin{equation}
\ln {\cal Z}[\Phi(t),\chi(t)]=\int
dt\,S\bigl(\dot{\Phi}(t),\chi(t)\bigr).\label{ZAdef}
\end{equation}
The path integral (\ref{concatenation}) can be taken in saddle-point
approximation, with the result
\begin{equation}
S_{A+B}(\dot{\Phi},\chi)
=S_{A}(\dot{\Phi}_{s},\chi_{s})+S_{B}(\dot{\Phi}-\dot{\Phi}_{s},
\chi-\chi_{s}).\label{saddlepoint}
\end{equation}
Here $\dot{\Phi}_{s}$ and $\chi_{s}$ stand for the (generally complex) values
of $\dot{\Phi}_{1}$ and $\chi_{1}$ at the saddle point (where the derivatives
with respect to these phases vanish).

Eqs.\ (\ref{ZAdef}) and (\ref{saddlepoint}) are quite general and now we apply
them to the specific circuit of Fig.\ \ref{circuit}. We assume that the
mesoscopic conductor $A$ (conductance $G$) is in series with a macroscopic
conductor $B$ with frequency dependent impedance $Z(\omega)$. (Any
electromagnetic environment with a Gaussian Keldysh action can be represented
by an external impedance.) The circuit is driven by a voltage source with
voltage $V_{0}$. To avoid Coulomb blockade effects \cite{ASI}, we assume that
$|Z(\omega)|\ll\hbar/e^{2}$ at frequencies $\hbar\omega\agt\max(eV,kT)$ (with
$T$ the temperature). The zero-frequency impedance $Z(0)\equiv Z_{0}\equiv
z_{0}h/e^{2}$ can have any value.

Both the voltage drop $V$ at the mesoscopic conductor and the current $I$
through the conductor fluctuate in time for finite $Z_{0}$, with averages
$\bar{I}=V_{0}G(1+Z_{0}G)^{-1}$, $\overline{V}=V_{0}(1+Z_{0}G)^{-1}$. Voltage
bias corresponds to $Z_{0}G\ll 1$ and current bias to $Z_{0}G\gg 1$, with
$I_{0}=V_{0}/Z_{0}$ the imposed current. There are three characteristic time
scales: $\hbar/\max(e\overline{V},kT)$, $e/\bar{I}$, and the $RC$-time of the
circuit. (The inverse of the $RC$-time is the frequency below which $Z\approx
Z_{0}$.) The low-frequency regime on which we concentrate is reached for
current and voltage fluctuations that are slow on any of these time scales.

We assume that the temperature is sufficiently low ($kT\ll e\overline{V}$) to
neglect thermal noise relative to shot noise. (Generalization to finite
temperature is straightforward, but will not be done here for simplicity.) The
low-temperature, low-frequency Keldysh action of the external impedance is
simply $S_{B}(\dot{\Phi},\chi) = i\chi\dot{\Phi}/2\pi z_{0}$, while the action
$S_{A}$ of the mesoscopic conductor is given by \cite{Lev93}
\begin{equation}
S_{A}(\dot{\Phi},\chi)=\frac{\dot{\Phi}}{2\pi}{\cal S}(i\chi),\;\;{\cal
S}(\xi)= \sum_{n=1}^{N}\ln\bigl[1+(e^{\xi}-1)T_{n}\bigr].\label{Sxi0}
\end{equation}
The $T_{n}$'s are the transmission eigenvalues, with
$\sum_{n}T_{n}=Gh/e^{2}\equiv g$ the dimensionless
conductance. (The formulas are written without including spin or other
degeneracies \cite{spin}.)

We seek the cumulant generating function of charge
\begin{equation}
{\cal F}(\xi)=\ln\left(\sum_{q=0}^{\infty}e^{q\xi}P(q)\right)=
\sum_{p=1}^{\infty}\langle\!\langle q^{p}\rangle\!\rangle
\frac{\xi^{p}}{p!},\label{Fxidef}
\end{equation}
where $\langle\!\langle q^{p}\rangle\!\rangle$ is the $p$-th cumulant of the
charge transferred during the time interval $\tau$. It is related to the
Keldysh action (\ref{saddlepoint}) by
\begin{equation}
{\cal F}(\xi)=\tau S_{A+B}(eV_{0}/\hbar,-i\xi).\label{calFdef}
\end{equation}
We also require the cumulant generating function of phase, ${\cal G}(\xi)$.
Since $V=V_{0}-Z_{0}I$ (in the absence of thermal noise from the external
impedance), it is related to ${\cal F}(\xi)$ by a change of variables (from $q$
to $\phi=\phi_{0}-qz_{0}$). The relation is
\begin{equation}
{\cal G}(\xi)=\sum_{p=1}^{\infty}\langle\!\langle
\phi^{p}\rangle\!\rangle\frac{\xi^{p}}{p!}=\phi_{0}\xi+{\cal
F}(-z_{0}\xi).\label{Gxidef}
\end{equation}

In the limit $Z_{0}\rightarrow 0$ of voltage bias the saddle point of the
Keldysh action is at $\dot{\Phi}_{s}=\dot{\Phi}$, $\chi_{s}=\chi$, and from
Eqs.\ (\ref{saddlepoint}), (\ref{Fxidef}), and (\ref{Gxidef}) one recovers the
results of Ref.\ \cite{Lev93}: The cumulant generating function ${\cal
F}_{0}(\xi)=\tau S_{A}(eV_{0}/\hbar,-i\xi)=\phi_{0}{\cal S}(\xi)$ and the
corresponding probability distribution
\begin{equation}
P_{\phi_{0}}(q)=\lim_{x\rightarrow 0}\frac{1}{q!}\frac{d^{q}}{dx^{q}}
\prod_{n=1}^{N}[1+(x-1)T_{n}]^{\phi_{0}}. \label{Pq0}
\end{equation}
The parameter $\phi_{0}=eV_{0}\tau/h$ is the number of attempted transmissions
per
channel, assumed to be an integer $\gg 1$. The first few cumulants are $\langle
q\rangle_{0}=\phi_{0}g$, $\langle\!\langle
q^{2}\rangle\!\rangle_{0}=\phi_{0}\sum_{n}T_{n}(1-T_{n})$, $\langle\!\langle
q^{3}\rangle\!\rangle_{0}=\phi_{0}\sum_{n}T_{n}(1-T_{n})(1-2T_{n})$. In the
single-channel case ($N=1$, $T_{1}\equiv\Gamma$) the distribution (\ref{Pq0})
has the binomial form (\ref{Pqresult}).

After these preparations we are now ready to generalize all of this to finite
$Z_{0}$, and in particular to derive the dual distribution of phase
(\ref{Pphiresult}) under current bias. The key equation that allows us to do
that follows directly from Eqs.\
(\ref{saddlepoint}) and (\ref{calFdef}):
\begin{equation}
{\cal F}(\xi)=\frac{\phi_{0}}{z_{0}}[\xi-\sigma(\xi)],\;\;
\sigma+z_{0}{\cal S}(\sigma)=\xi.\label{Fxi}
\end{equation}
The implicit function $\sigma(\xi)$ (which determines the saddle point of the
Keldysh action) provides the cumulant generating function of charge ${\cal F}$
for arbitrary series resistance $z_{0}=(e^{2}/h)Z_{0}$. One readily checks that
${\cal F}(\xi)\rightarrow\phi_{0}{\cal S}(\xi)$ in the limit $z_{0}\rightarrow
0$, as it should.

By expanding Eq.\ (\ref{Fxi}) in powers of $\xi$ we obtain a relation between
the cumulants $\langle\!\langle q^{p}\rangle\!\rangle$ of charge at $Z_{0}\neq
0$ and the cumulants $\langle\!\langle q^{p}\rangle\!\rangle_{0}$ at $Z_{0}=0$.
For example, to linear order we find $\langle q\rangle=(1+z_{0}g)^{-1}\langle
q\rangle_{0}$, which amounts to the obvious statement that the mean current
$\bar{I}$ is rescaled by a factor $1+Z_{0}G$ as a result of the series
resistance. The Langevin approach discussed in the introduction predicts that
the same rescaling applies to the fluctuations. Indeed, to second order we find
$\langle\!\langle q^{2}\rangle\!\rangle=(1+z_{0}g)^{-3}\langle\!\langle
q^{2}\rangle\!\rangle_{0}$, in agreement with Ref.\ \cite{Bla00}.

However, if we go to higher cumulants we find that other terms appear, which
can not be incorporated by any rescaling. For example, Eq.\ (\ref{Fxi}) gives
for the third cumulant
\begin{equation}
\langle\!\langle q^{3}\rangle\!\rangle=\frac{\langle\!\langle
q^{3}\rangle\!\rangle_{0}} {(1+z_{0}g)^{4}}- \frac{3z_{0}g}
{(1+z_{0}g)^{5}}\frac{\bigl(\langle\!\langle
q^{2}\rangle\!\rangle_{0}\bigr)^{2}}{\langle
q\rangle_{0}}.\label{thirdcumulant}
\end{equation}
The first term on the the right-hand-side has the expected scaling form, but
the second term does not. This is generic for $p\geq 3$:  $\langle\!\langle
q^{p}\rangle\!\rangle=(1+z_{0}g)^{-p-1}\langle\!\langle q^{p}\rangle\!\rangle$
plus a non-linear (rational) function of lower cumulants \cite{note3}. All
terms are of the same order of magnitude in $z_{0}g$, so one can not neglect
the non-linear terms.

Turning now to the limit $z_{0}g\rightarrow\infty$ of current bias, we see from
Eq.\ (\ref{Fxi}) that ${\cal F}\rightarrow {\cal F}_{\infty}$ with
\begin{equation}
{\cal F}_{\infty}(\xi)=q_{0}\xi-q_{0}{\cal S}^{\rm
inv}(\xi/z_{0})\label{Finfty}
\end{equation}
defined in terms of the functional inverse ${\cal S}^{\rm inv}$ of ${\cal S}$.
The parameter
$q_{0}=\phi_{0}/z_{0}=I_{0}\tau/e$ (assumed to be an integer $\gg 1$) is the
number of charges transferred by the imposed current $I_{0}$ in the detection
time $\tau$. Transforming from charge to phase variables by means of Eq.\
(\ref{Gxidef}), we find that ${\cal G}\rightarrow {\cal G}_{\infty}$ with
\begin{equation}
{\cal G}_{\infty}(\xi)=-q_{0}{\cal S}^{\rm inv}(-\xi).\label{Ginfty}
\end{equation}
In the single-channel case Eq.\ (\ref{Ginfty}) reduces to ${\cal
G}_{\infty}(\xi)=-q_{0}\ln[1+\Gamma^{-1}(e^{-\xi}-1)]$, corresponding to the
Pascal distribution (\ref{Pphiresult}). The first three cumulants are
$\langle\phi\rangle=q_{0}/\Gamma$,
$\langle\!\langle\phi^{2}\rangle\!\rangle=(q_{0}/\Gamma^{2})(1-\Gamma)$,
$\langle\!\langle\phi^{3}\rangle\!\rangle=
(q_{0}/\Gamma^{3})(1-\Gamma)(2-\Gamma)$.

For the general multi-channel case a simple expression for
$P_{q_{0}}(\phi)$ can be obtained in the ballistic limit (all $T_{n}$'s close
to 1) and in the tunneling limit (all $T_{n}$'s close to 0). In the ballistic
limit one has ${\cal G}_{\infty}(\xi)=q_{0}\xi/N+q_{0}(N-g)(e^{\xi/N}-1)$,
corresponding to a Poisson distribution in the discrete variable
$N\phi-q_{0}=0,1,2,\ldots$. In the tunneling limit ${\cal
G}_{\infty}(\xi)=-q_{0}\ln(1-\xi/g)$, corresponding to a chi-square
distribution $P_{q_0}(\phi)\propto \phi^{q_{0}-1}e^{-g\phi}$ in the
continuous variable $\phi>0$.  In contrast, the charge distribution
$P_{\phi_{0}}(q)$
is Poissonian both in the tunneling limit (in the variable $q$) and in the
ballistic limit (in the variable $N\phi_{0}-q$).

For large $q_{0}$ and $\phi$, when the discreteness of these variables can be
ignored, we may calculate $P_{q_{0}}(\phi)$ from ${\cal G}_{\infty}(\xi)$ in
saddle-point approximation. If we also calculate $P_{\phi_{0}}(q)$ from ${\cal
F}_{0}(\xi)$ in the same approximation (valid for large $\phi_{0}$ and $q$), we
find that the two distributions have a remarkably similar form:
\begin{eqnarray}
P_{\phi_{0}}(q)&=&N_{\phi_{0}}(q)\exp[\tau\Sigma(2\pi\phi_{0}/\tau,q/\tau)],
\label{Pphi0q}\\
P_{q_{0}}(\phi)&=&N_{q_{0}}(\phi)\exp[\tau\Sigma(2\pi\phi/\tau,q_0/\tau)]
\label{Pq0phi}.
\end{eqnarray}
The same exponential function
\begin{equation}
\Sigma(x,y)=S_{A}(x,-i\xi_{s})-y\xi_{s}\label{Sigmadef}
\end{equation}
appears in both distributions (with $\xi_{s}$ the location of the saddle
point). The pre-exponential functions $N_{\phi_{0}}$ and $N_{q_{0}}$ are
different, determined by the Gaussian integration around the saddle point.
Since these two functions vary only algebraically, rather than exponentially,
we conclude that Eq.\ (\ref{PCP}) holds with the remainder ${\cal
O}(1)=\ln(q/\phi)$ obtained by evaluating $\ln[2\pi(\partial^{2}\Sigma/\partial
x^{2})^{1/2}(\partial^{2}\Sigma/\partial y^{2})^{-1/2}]$ at $x=2\pi\phi/\tau$,
$y=q/\tau$.

\begin{figure}
\includegraphics[width=7cm]{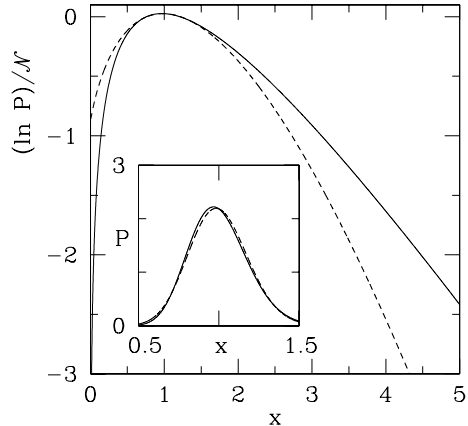}
\caption{Comparison of the distributions of charge (dashed curve, with
$x=q/\langle q\rangle$) and of phase (solid curve, with
$x=\phi/\langle\phi\rangle$), calculated from Eqs.\
(\protect\ref{PchargelargeN}) and (\protect\ref{PphaselargeN}) for ${\cal
N}=q_{0}=\phi_{0}\Gamma=30$ transferred charges in the tunneling limit
$\Gamma\ll 1$. The main plot emphasizes the non-Gaussian tails on a
semi-logarithmic scale, the inset shows on a linear scale that the Gaussian
body of the distributions coincides.}  \label{logP}
\end{figure}

The distributions of charge and phase are compared graphically in Fig.\
\ref{logP}, in the tunneling limit $\Gamma\ll 1$. We use the rescaled variable
$x=q/\langle q\rangle$ for the charge and $x=\phi/\langle\phi\rangle$ for the
phase, and take the same mean number ${\cal N}=q_{0}=\phi_{0}\Gamma$ of
transferred charges in both cases. We plot the asymptotic large-${\cal N}$ form
of the distributions,
\begin{eqnarray}
P_{\rm charge}(x)&=&({\cal N}/2\pi)^{1/2}x^{-1/2}e^{{\cal N}(x-1-x\ln
x)},\label{PchargelargeN}\\
P_{\rm phase}(x)&=&({\cal N}/2\pi)^{1/2}x^{-1} e^{{\cal N}(1-x+\ln
x)},\label{PphaselargeN}
\end{eqnarray}
corresponding to the Poisson and chi-square distribution, respectively. Since
the first two moments are the same, the difference appears in the non-Gaussian
tails. The difference should be readily visible as a factor of two in a
measurement of the third cumulant: $\langle\!\langle
x^{3}\rangle\!\rangle={\cal N}^{-2}$ for the charge and $\langle\!\langle
x^{3}\rangle\!\rangle=2{\cal N}^{-2}$ for the phase.

In summary, we have demonstrated theoretically that electrical noise becomes
intrinsically different when the conductor is current biased rather than
voltage biased. While the second cumulants can be related by a rescaling with
the conductance, the third and higher cumulants can not. Experiments are
typically carried out in an intermediate regime, and for that purpose we have
derived explicit expressions [such as Eq.\ (\ref{thirdcumulant})] for the
crossover from voltage to current bias. The non-linear term that breaks the
rescaling is of the same order of magnitude as the linear term, so it can not
be neglected. From a fundamental point of view, the limit of full current bias
is of particular interest. The counterpart of the celebrated binomial
distribution of transferred charge \cite{Lev93} turns out to be the Pascal
distribution of phase increments.

This work was supported by the Dutch Science Foundation NWO/FOM.


\begin{thebibliography}{99}
\bibitem{note1} In particular, the noise spectral densities
$P_{I}=\int_{-\infty}^{\infty}\!dt\, \langle\delta I(0)\delta I(t)\rangle$,
$P_{V}=\int_{-\infty}^{\infty}\!dt\, \langle\delta V(0)\delta V(t)\rangle$ of
current and voltage are given, respectively, by the variances of charge and
phase in the limit of infinite detection time:
$P_{I}=\lim_{\tau\rightarrow\infty}\tau^{-1} {\rm Var}\,Q$,
$P_{V}=(\hbar/e)^{2}\lim_{\tau\rightarrow\infty}\tau^{-1} {\rm Var}\,\Phi$.
Higher cumulants of $Q$ and $\Phi$ are similarly related to higher order
correlators of $I$ and $V$.
\bibitem{spin} Spin degeneracy can be accounted for by the substitution
$\phi_{0}\rightarrow 2\phi_{0}$ in Eq.\ (\protect\ref{Pqresult}) and
$\phi\rightarrow 2\phi$ in Eq.\ (\protect\ref{Pphiresult}), with $\phi$
half-integer.
\bibitem{note2} The Pascal distribution $P(m)={m-1\choose
M-1}\Gamma^{M}(1-\Gamma)^{m-M}$ is also called the ``binomial waiting-time
distribution'', since it gives the probability of the number $m$ of independent
trials (with success probability $\Gamma$) that one has to wait until the
$M$-th success. It is related to the negative-binomial distribution
$P(n)={n+M-1\choose n}\Gamma^{M}(1-\Gamma)^{n}$ by the displacement $n=m-M$.
\bibitem{note3} We record the result for the fourth cumulant, obtained by
expansion of Eq.\ (\protect\ref{Fxi}) to order $\xi^{4}$: $\langle\!\langle
q^{4}\rangle\!\rangle=(1+z_{0}g)^{-5}\mu_{4}-10z_{0}
g(1+z_{0}g)^{-6}\mu_{2}\mu_{3}/\mu_{1}+15(z_{0}g)^{2}(1+z_{0}g)^{-7}
\mu_{2}^{3}/\mu_{1}^{2}$, where we have abbreviated
$\langle\!\langle q^{p}\rangle\!\rangle_{0}=\mu_{p}$.
\end{thebibliography}
\end{document}